\documentclass{PoS}
\let\ifpdf\relax
\pdfoutput=1
\usepackage{cite}
\usepackage{amsmath,amsbsy,amssymb} 
\usepackage{graphicx}
\usepackage{epsfig}
\usepackage{fixmath}
\usepackage{caption}
\usepackage{subcaption}
\usepackage{slashed}
\graphicspath{plots/} 

\newcommand\new{\newcommand}         

\def\be{\begin{equation}}   
\def\ee{\end{equation}}
\def\bea{\begin{eqnarray}}  
\def\eea{\end{eqnarray}}

\def\MSbar{$\overline{{\rm MS}}$}

\def\gev{\mathrm{\:GeV}}
\def\mrm{\mathrm}
\new{\as}[1]      {{\ifmmode\alpha^{#1}_s
                    \else$\alpha^{#1}_s$\fi}}
\new{\lqcd}       {{\ifmmode\Lambda_{\mathrm{ QCD}}
                    \else $\Lambda_{\mathrm{ QCD}}$\fi}}

\new{\mpi}{Max-Planck-Institut f\"ur Physik, F\"ohringer Ring 6, 80805 M\"unchen, Germany}

\title{NLO QCD corrections to $\mathbold{W^+W^-b\bar{b}}$ production 
and top quark observables}

\ShortTitle{NLO QCD corrections to WWbb production and top quark observables}

\author{\speaker{Johannes~Schlenk}, Gudrun~Heinrich, Jan~Winter\\
Max Planck Institute for Physics, F\"ohringer Ring 6, 80805 Munich, Germany\\
jschlenk@mpp.mpg.de, gudrun@mpp.mpg.de, jwinter@mpp.mpg.de} 

\abstract{
We present the NLO QCD corrections to the production of a 
$W^+W^-$ pair and two $b$-jets, including the leptonic decays of the $W$ bosons.
Contributions from singly resonant and non-resonant 
top quarks are fully taken into account.
We also discuss observables relevant for top quark mass measurements   
and  $t\bar t$ asymmetries.}

\FullConference{11th International Symposium on Radiative Corrections (Applications of Quantum Field Theory to Phenomenology) \\
		 22-27 September 2013\\
		 Lumley Castle Hotel, Durham, UK}

\begin{document}

\section{Introduction}

Precision measurements of observables related to top quarks are of
primary importance at the LHC and future colliders, to obtain more
information on the Higgs sector as well as indirect hints of physics
beyond the Standard Model.
However, quantities like the top quark mass or the forward--backward
asymmetry are unphysical observables, in the sense that they have to
be reconstructed from the top quark decay products. It therefore is a
non-trivial task to match the theory predictions with the quantities
reconstructed from the experimental measurements.
To have a precise description from the theory side, 
the predictions need to go beyond the simple approximation 
of factorising top quark production and decay.
For example, finite width effects and non-factorising contributions to
observables based on $W$ boson decay products and $b$-jets can have a
non-negligible impact on mass measurements.

The next-to-leading order QCD corrections to top quark pair
production~\cite{Nason:1987xz,Nason:1989zy,Beenakker:1990maa,Mangano:1991jk,Frixione:1995fj}
recently have been enhanced by the NNLO corrections to the total
cross section~\cite{Czakon:2013goa}.
The NLO electroweak corrections are also known~\cite{Beenakker:1993yr}. 
These calculations treat the top quarks as stable on-shell particles.
If included, their decays have been computed in the narrow width
approximation (NWA), where production and decay decouple. NLO
calculations in the NWA were further improved
in \cite{Melnikov:2009dn,Melnikov:2011ai,Biswas:2010sa},  
where spin correlations between top quark production and decay have
been taken into account.

The full process $pp(p\bar p)\rightarrow W^+W^-b\bar{b}$ was recently
calculated at NLO QCD in~\cite{Denner:2010jp,Denner:2012yc,Bevilacqua:2010qb,Heinrich:2013qaa} 
for massless $b$-quarks, and in~\cite{Frederix:2013gra,Cascioli:2013wga} in the 4-flavour scheme, 
i.e. for massive $b$-quarks.
It represents a $2\rightarrow 4$ process which is of much higher
complexity than the factorised process discussed above. In this talk
we present the NLO QCD corrections to
$pp(p\bar p)\rightarrow W^+W^-b\bar{b} \rightarrow
(e^+ \nu_e)(\mu^- \bar{\nu}_{\mu})b\bar{b}$ in the 5-flavour scheme, 
including singly-resonant and non-resonant contributions, 
corresponding to Feynman diagrams containing only one or no top quark
propagator that can go on-shell. 
The impact of non-resonant $W$ boson
contributions has been studied in \cite{Denner:2012yc} and was found
to be small. Therefore, non-resonant contributions from $W$ bosons are
neglected in our calculation. On the other hand, in contrast to the
calculations in \cite{Denner:2010jp,Denner:2012yc,Bevilacqua:2010qb}, 
contributions from massless $b$ quarks in the initial state are 
included in our calculation. 
For more details we refer to~\cite{Heinrich:2013qaa}.


\section{Calculational Setup}
\label{sec:calc}

The virtual corrections were calculated by the automated one-loop
generator \textsc{GoSam}~\cite{Cullen:2011ac}.\footnote{\textsc{GoSam}~is
publicly available at \texttt{http://gosam.hepforge.org/}.}
The program combines cut-based integrand reduction 
techniques~\cite{Ossola:2006us,Ellis:2007br,Mastrolia:2010nb,Heinrich:2010ax,vanDeurzen:2013pja,vanDeurzen:2013saa} 
with improved tensor reduction
methods~\cite{Binoth:2005ff,Binoth:2008uq,Cullen:2011kv}.
The basis integrals are taken from \textsc{golem95C}~\cite{Cullen:2011kv,Guillet:2013msa}
or \textsc{OneLOop}~\cite{vanHameren:2010cp}.

For the real radiation and the NLO infrared subtraction terms 
as well as for the Monte Carlo integration, the event generator 
\textsc{Sherpa}~\cite{Gleisberg:2007md,Gleisberg:2008ta} has been used, 
interfaced to \textsc{GoSam} via the Binoth--Les--Houches accord
(BLHA)~\cite{Binoth:2010xt,Alioli:2013nda}. For more applications of \textsc{GoSam}, we refer to the contributions \cite{gosamtalks1,gosamtalks2}.

%

As mentioned in the introduction, our calculation includes
singly-resonant and non-resonant contributions. To take the top quark
decay width into account in a gauge invariant way, 
the complex mass scheme~\cite{Denner:2006ic} is used. 
In our case, this amounts to replacing the top quark mass everywhere
by a complex parameter $\mu_t$ according to $\mu_t^2=m_t^2-im_t\upGamma_t$.


The correctness of our virtual amplitude has been checked by comparing
it with the results of \cite{Denner:2012yc}.
The real radiation part was checked by calculating the cross section
for different values of the dipole $\alpha$-parameter~\cite{Nagy:1998bb}, 
i.e.~for $\alpha_{\mrm{dip}}=\{0.1,\,0.05,\,0.01\}$, and the results were
found to be in agreement within the statistical uncertainty. 

\begin{figure}[t!]
\centering
        \begin{subfigure}[b]{0.37\textwidth}
                \centering
\includegraphics[width=1\textwidth]{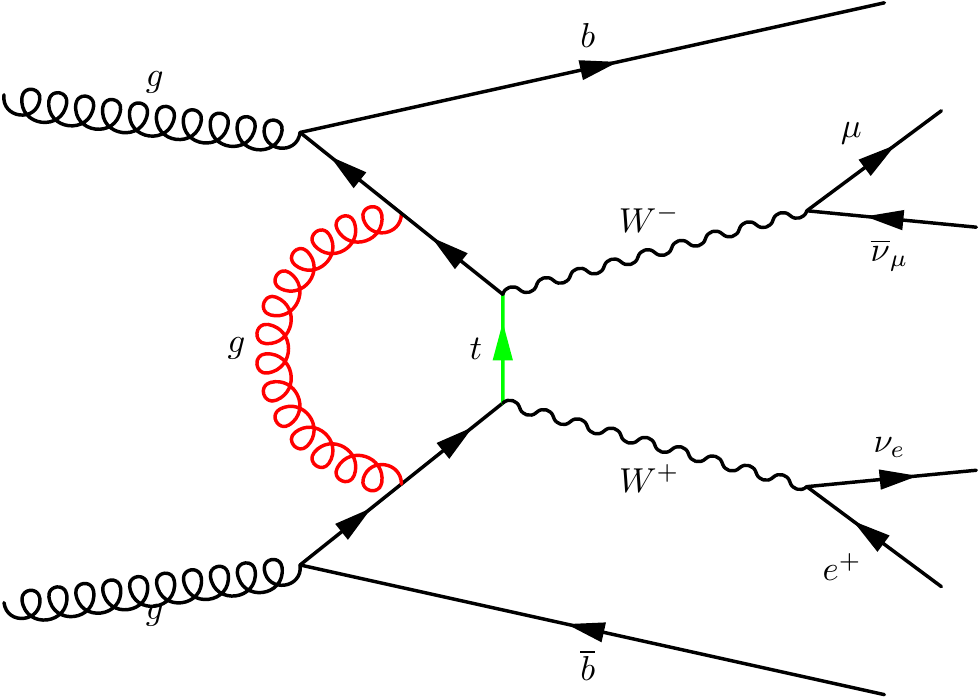}
        \end{subfigure}
        ~ 
        \begin{subfigure}[b]{0.37\textwidth}
                \centering
\includegraphics[width=1\textwidth]{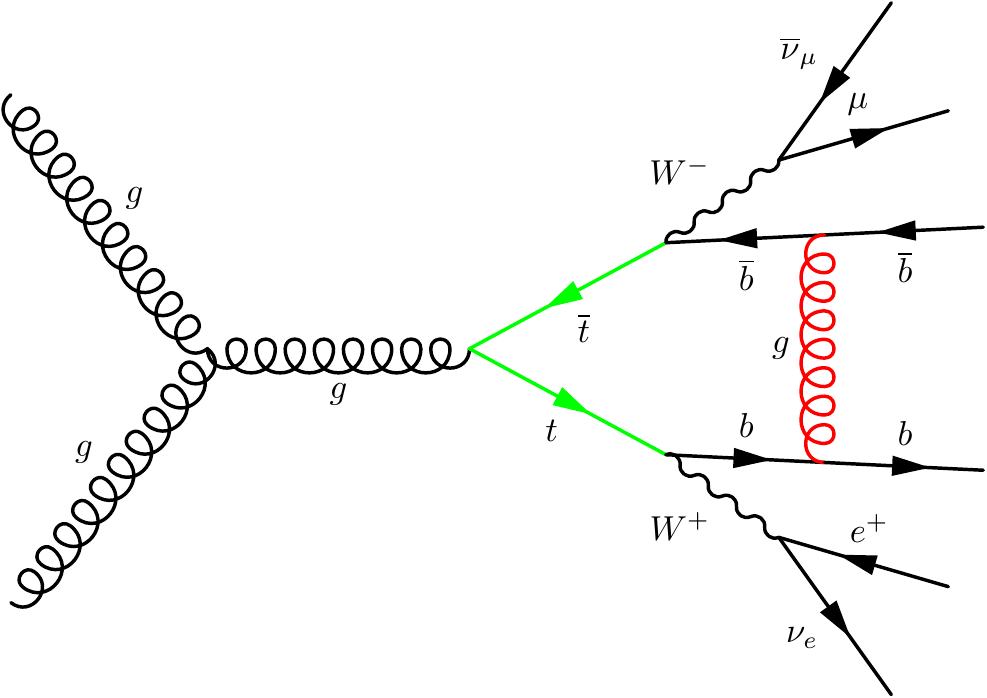}
        \end{subfigure}
\caption{Example of a non-resonant diagram and a non-factorizable
virtual contribution.}
\label{fig:diagloop}
\end{figure} 

At NLO, the NWA neglects non-resonant diagrams and radiative
corrections that connect production and decay or both decays.
Two example Feynman diagrams contributing to the virtual corrections, 
which are not present in the NWA, are given in Figure \ref{fig:diagloop}.
One expects 
that the contributions neglected in the NWA are suppressed by powers
of $\upGamma_t/m_t \lesssim 1\%$. While this is true for sufficiently
inclusive observables, the corrections can be much larger for
observables such as the invariant mass of the charged lepton plus
$b$-jet, $m_{lb}$~\cite{AlcarazMaestre:2012vp}.

\section{Phenomenological Results}
\label{sec:pheno}

\subsection{General input parameters and LHC cross sections}

For the (N)LO calculations, the MSTW2008(N)LO parton
distributions~\cite{Martin:2009iq} were used, taking the strong
coupling constant $\alpha_S$ and its running as provided by these
PDFs. For the electroweak parameters, we employ 
$G_{\mu} =1.16637\cdot10^{-5}\gev^{-2}$,
$M_{W} =80.399\gev$, $\upGamma_W = 2.0997\gev$,
$M_{Z} =91.1876\gev$, $\upGamma_Z = 2.5097\gev$.
Diagrams involving a Higgs boson propagator are neglected in the
entire calculation owing to their small contribution. All quarks other
than the top quark are taken to be massless. For the top quark mass
and widths at leading and next-to-leading order, we use
$m_t=172.0\gev$,
$\upGamma_t^{\mrm{LO}} = 1.4426\gev$ and
$\upGamma_t^{\mrm{NLO}} =1.3167\gev$, respectively.

Results are presented for the LHC at $7\mrm{\:TeV}$ centre-of-mass energy.
All final state partons are clustered into jets with a geometric separation
$\Delta R =\sqrt{\Delta \phi^2 + \Delta \eta^2} > 0.5$
using the anti-$k_T$ jet algorithm
\cite{Cacciari:2005hq,Cacciari:2008gp} implemented in \textsc{FastJet}
\cite{Cacciari:2011ma}. Each event must contain at least two $b$-jets
obeying the conditions $p_{T,b}>30\gev$ and $|\eta_{b}|<2.5$. The
kinematic requirements on the charged leptons and the missing energy
are: $p_{T,l}>20\gev$, $|\eta_{l}|<2.5$ and
$\slashed{p}_T>20\gev$.\footnote{Here, we determine the missing energy
  from the transverse vector sum of the neutrinos.}
The renormalisation and the factorisation scales are set to $\hat H_T/2$.
The variable $\hat H_T$ is defined as $\hat H_T=\sum_{j}p_{T,j}\;$, 
where the sum goes over {\em all} final-state partons, including
leptons.
The particular choice of $\hat H_T/2$ for the central scale is made
because of the observation that the difference between the LO and NLO
cross sections as well as the uncertainty introduced by the scale
variation remain relatively small.


\begin{figure}[t!]
\centering
        \begin{subfigure}[b]{0.45\textwidth}
                \centering
\includegraphics[width=1\textwidth]{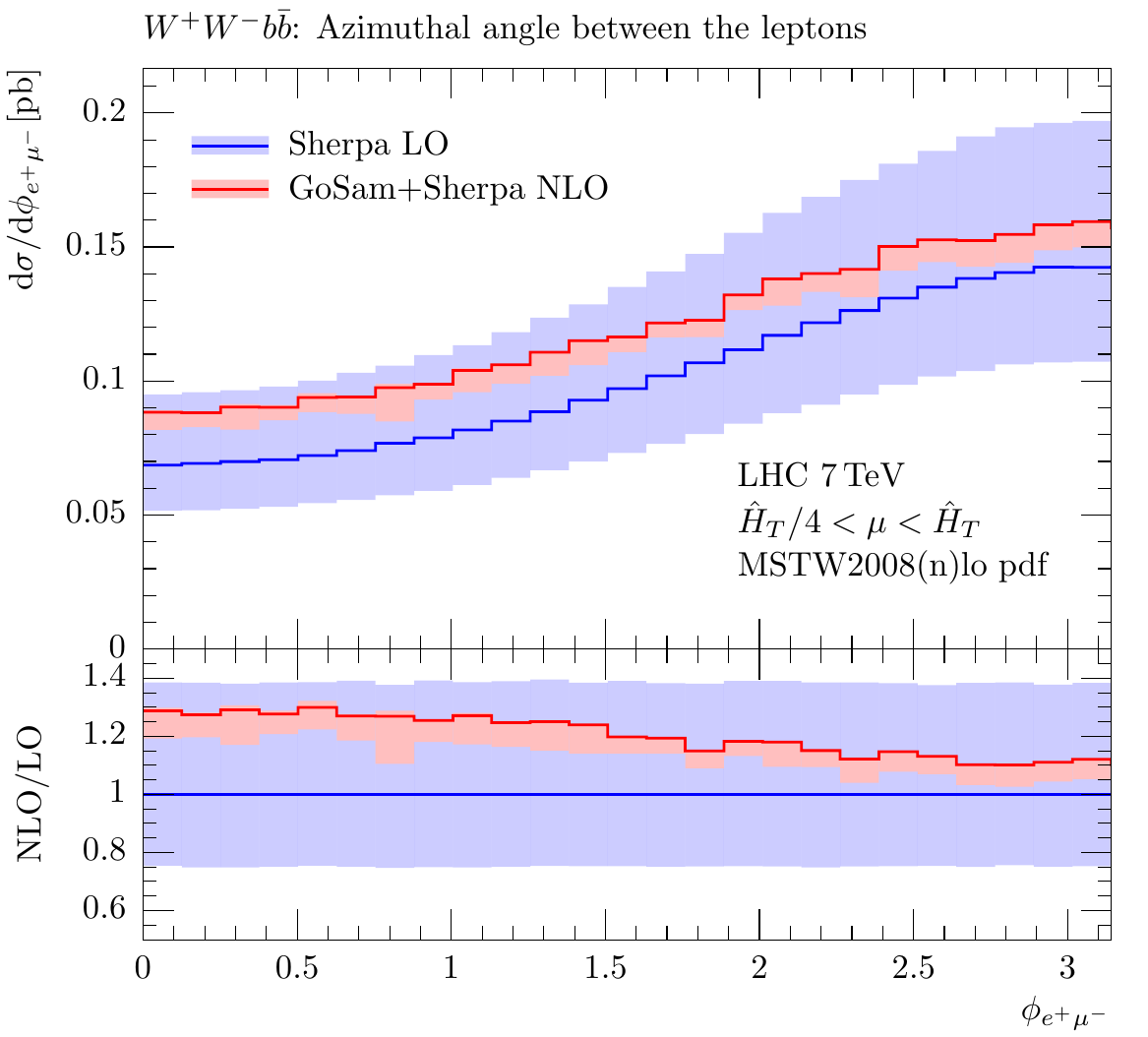}
        \end{subfigure}
        \begin{subfigure}[b]{0.45\textwidth}
                \centering
\includegraphics[width=1\textwidth]{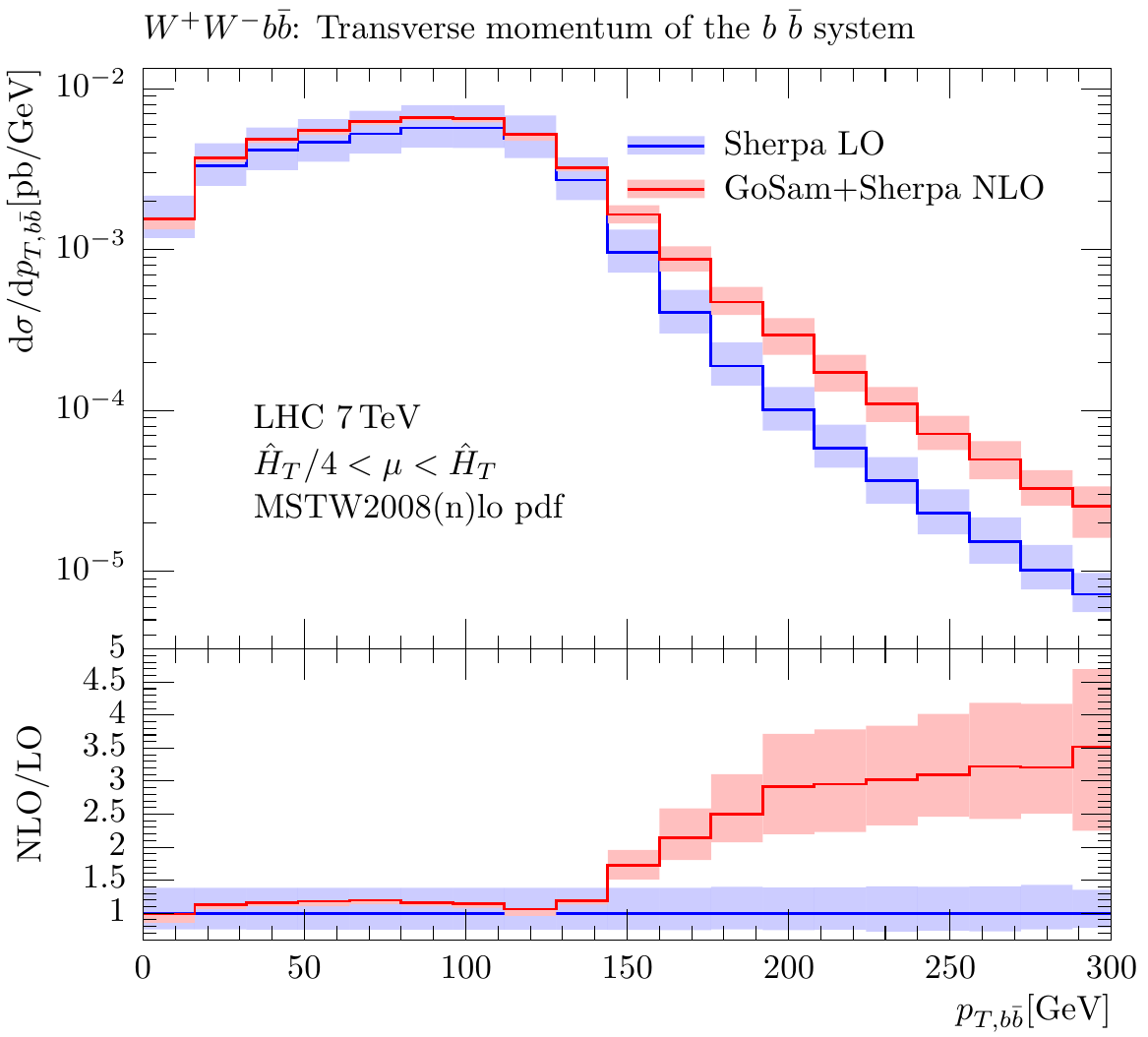}
        \end{subfigure}
\caption{The azimuthal angle distributions of the two charged leptons
  (left) and the transverse momentum distributions of the  $b\bar{b}$
  system (right). The bands around the distributions are obtained by
  varying $\mu_\mrm{R,F}$ simultaneously by a factor of two around the
  central scale $\hat H_T/2$.}
\label{fig:angle}
\end{figure}

The cross sections obtained with the parameters given above are
\begin{equation}
\begin{aligned}
\sigma_{\mrm{LO}}\,\mrm{[fb]} &\;=\; 638.4^{+38.5\%}_{-24.8\%}\,\mrm{(scale)}
\;\pm\;0.03\%\,\mrm{(stat)}\\
\sigma_{\mrm{NLO}}\,\mrm{[fb]} &\;=\; 758.5^{-2.5\%}_{-5.3\%}\,\mrm{(scale)}
\;\pm\;0.2\%\,\mrm{(stat)}~.
\end{aligned}
\label{eq:xs}
\end{equation}
Figure \ref{fig:angle}, to the left, shows the distribution of the
azimuthal angle between the two charged leptons, $\phi_{e^+\mu^-}$,
stemming from the $W$ boson decays. This angle plays an important role
for the measurement of spin correlations.
We observe a substantial reduction of the scale uncertainty at NLO.
The $\phi_{e^+\mu^-}$ distribution receives large NLO corrections in regions 
where the separation between the two leptons is small, with a
variation of the $K$-factor of $\sim20\%$.
Figure \ref{fig:angle}, to the right, shows the transverse momentum of
the vector sum of the two $b$-jet momenta. While for low $p_T$, one
again finds $K$-factors of ${\cal O}(1.2)$, this observable receives
large NLO corrections above $p_T\simeq150\gev$. The reason for this
$K$-factor increase up to $3$ in the tail of $p_{T,bb}$ lies in the
generation of real radition at NLO. At LO, the $t\bar{t}$ pair has
zero transverse momentum, which leads to a suppression of
$b\bar{b}$ pairs with high transverse momentum. At NLO, it however can
obtain transverse momentum by recoiling against the real radiation.

\subsection{Reconstruction of the top quark mass}

The mass $m_t$ of the top quarks can be reconstructed by measuring
kinematical distributions of their decay products.
As $m_t$ is not a physical observable, it is scheme dependent. 
The most commonly used mass definitions are the pole mass and the 
\MSbar{} mass. The different masses are related by a perturbative series, 
see e.g.~\cite{Chetyrkin:1999ys,Melnikov:2000qh}.
There are several issues, which render a precise top quark mass
determination at hadron colliders difficult, such as the dependence on
the definition of the top quark mass, $b$-mass and non-perturbative
effects in the Monte Carlo modeling, finite width effects and bound
state effects. For a recent review, we refer to
\cite{Agashe:2013hma,Juste:2013dsa}.

\begin{figure}[t!]
\centering
        \begin{subfigure}[b]{0.49\textwidth}
                \centering
\includegraphics[width=1\textwidth]{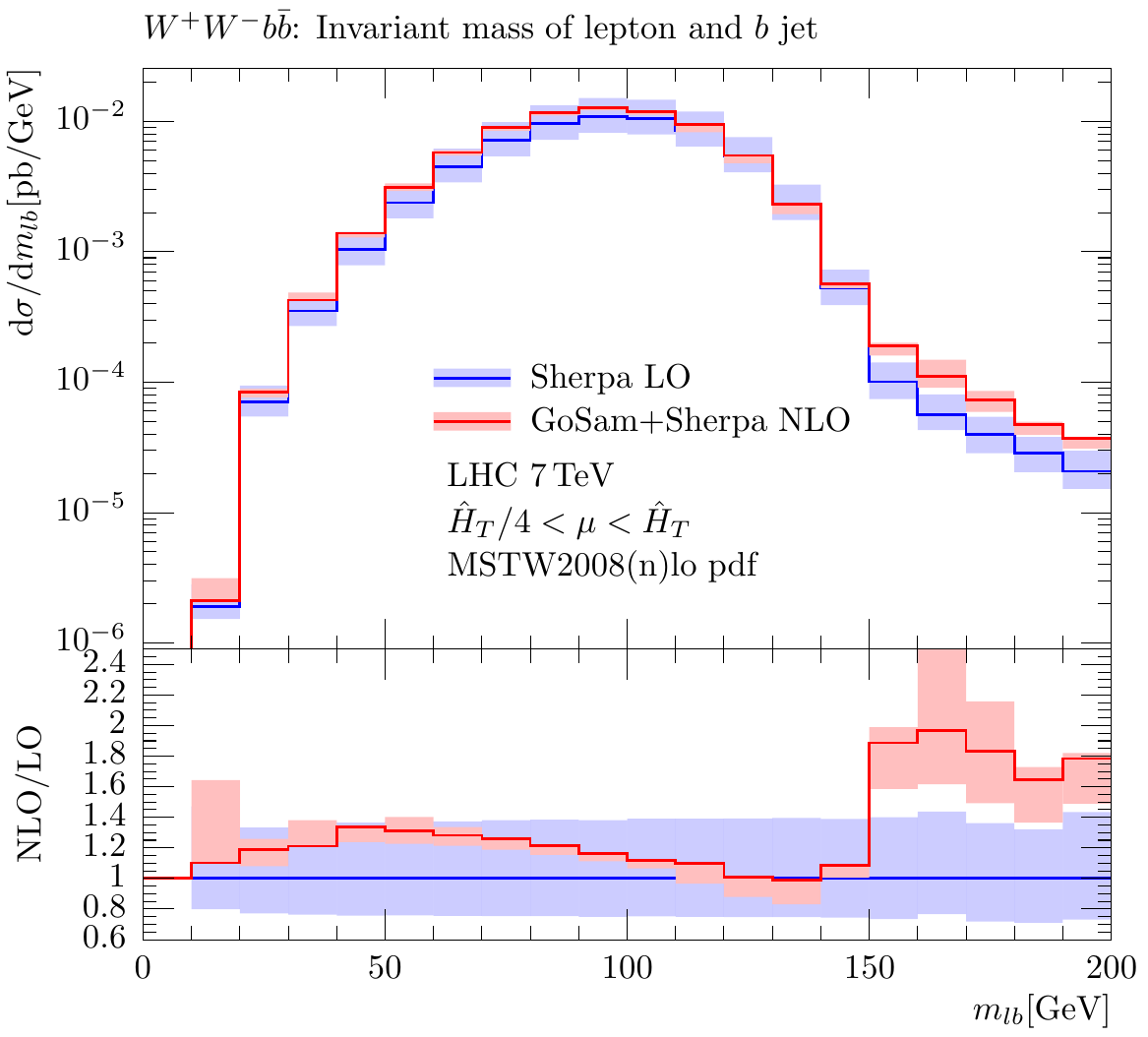}
        \end{subfigure}
        \begin{subfigure}[b]{0.49\textwidth}
                \centering
\includegraphics[width=1\textwidth]{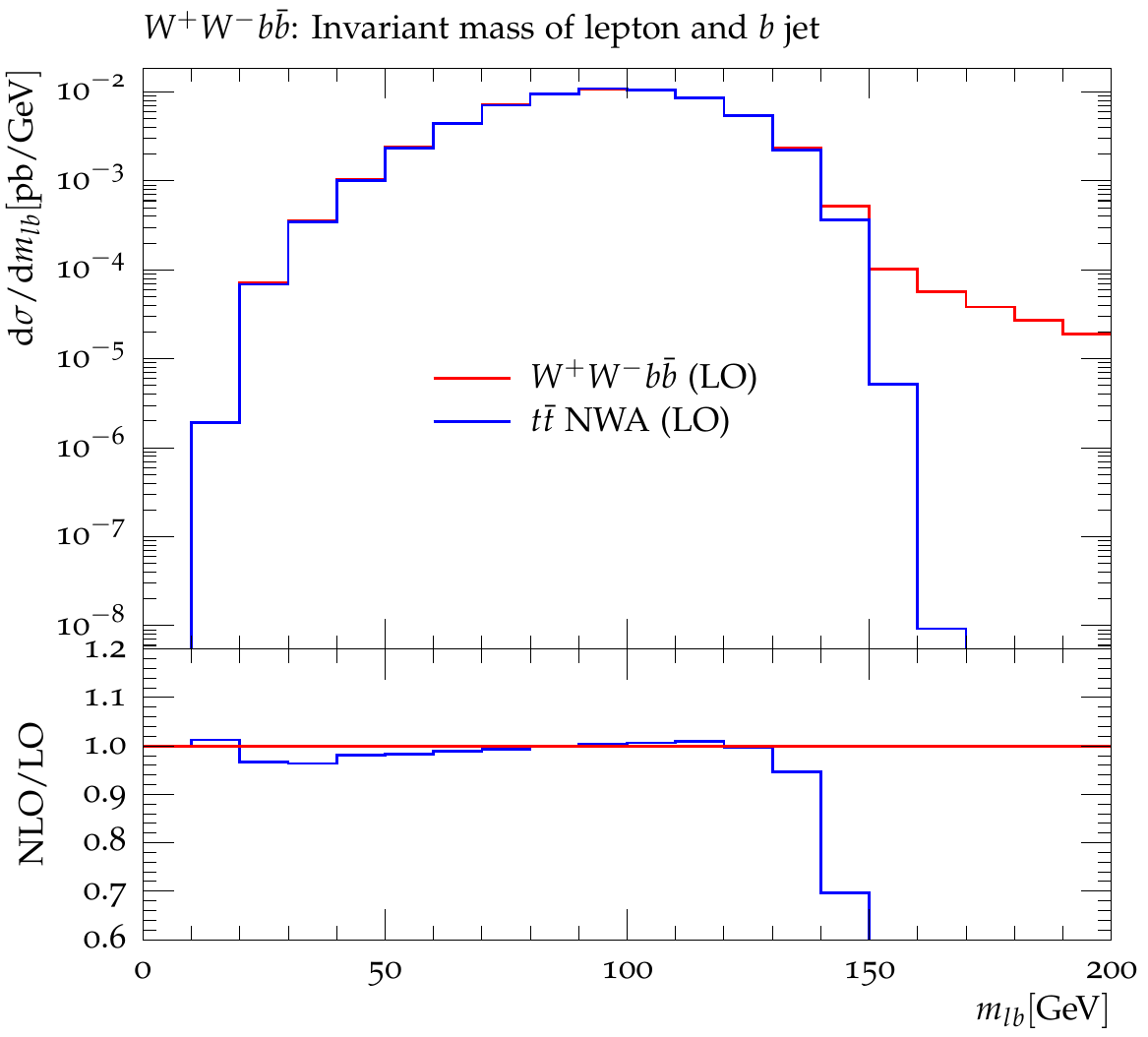}
        \end{subfigure}
\caption{The invariant mass $m_{lb}$ distribution at NLO and LO (left)
  and the comparison to the LO prediction in the NWA (right).}
\label{fig:mlbnlo}
\end{figure}

Here we focus on the reconstruction of the top quark mass from the
distribution of the invariant mass of a charged lepton and a $b$-jet,
$m_{lb}=(p_l+p_b)^2$, shown in Figure~\ref{fig:mlbnlo}.
Top quark mass measurements based on this observable have been
performed e.g.~in \cite{ATLAS:2012aj,ATLAS-CONF-2013-077,Chatrchyan:2012ea,%
Aaltonen:2011dr,Abazov:2012rp}.
For our calculation of the $m_{lb}$ distribution, we use the ATLAS
cuts of \cite{ATLAS-CONF-2013-077}. We require exactly two oppositely
charged leptons (electrons with $p_{T,e}>25\gev$ and muons with
$p_{T,\mu}>20\gev$) with $|\eta_l|<2.5$ and two $b$-jets with $p_{T,b}>25\gev$,
$|\eta_b|<2.5$ and $\Delta R>0.4$. The leptons have to be isolated
from the jets according to $\Delta R_{l,\mrm{jet}}>0.4$. Furthermore,
$H_T$ defined as the sum over the transverse momenta of charged
leptons and jets has to be larger than $130\gev$.

One complication arises from the fact that there are two top quarks
and therefore two possible $m_{lb}$ values in each event per charged
lepton. Since the charge of the bottom quark initiating the jet has
not been reconstructed experimentally, one needs a criterion to assign
$b$-jets to the ``correct'' lepton. Different strategies for the
assignment were tested based on Monte Carlo studies. It turned out
that the requirement of minimising the sum of both $m_{lb}$ led to the
best criterion in identifying the optimal $lb$ pairings, with a
recombination efficiency of about 77 percent~\cite{ATLAS-CONF-2013-077}.
The histogram to the right in Figure~\ref{fig:mlbnlo} shows a LO
comparison between the full $W^+W^-b\bar{b}$ calculation and the NWA.
In the NWA, $m_{lb}$ ideally has a sharp cut-off at $\sqrt{m^2_t-m^2_W}$,
which corresponds to the limit of vanishing neutrino momentum.
The non-resonant contributions included in the full calculation lead to a 
tail in the distribution. This shows that, while the integrated cross
section receives only small corrections, non-resonant contributions can
become sizeable for some distributions.

\subsection{Top quark asymmetries}
\label{sec:asym}

At $p\bar p$ colliders, the top quark forward--backward asymmetry
$A_{t\bar{t}}^{FB}$ is defined as
\begin{equation}
A_{t\bar{t}}^{FB}\;=\;\frac{\sigma\left(\Delta y>0\right) -  \sigma\left(\Delta y<0\right)}{ \sigma\left(\Delta y>0\right) + \sigma\left(\Delta y<0\right)}
\label{eq:fbast}
\end{equation}
using the rapidity difference $\Delta y = y_t -y_{\bar{t}}$. For
$t\bar t$ production calculated at LO in QCD, this asymmetry
vanishes. The first non-zero contribution to $A_{t\bar{t}}^{FB}$
appears at NLO. Measurements of $A_{t\bar{t}}^{FB}$ at the Tevatron
\cite{Abazov:2011rq,Aaltonen:2011kc}
give a significantly larger value than the Standard Model prediction
\cite{Kuhn:2011ri,Hollik:2011ps,Bernreuther:2012sx}.

As the top quarks have to be reconstructed from their decay products, 
the experimental results for $A_{t\bar{t}}^{FB}$ depend on the reconstruction method.
To avoid a bias in this procedure, one can also define a leptonic asymmetry 
\begin{equation}
A_{l^+l^-}^{FB}\;=\;\frac{\sigma\left(\Delta \eta>0\right) -  \sigma\left(\Delta \eta<0\right)}{ \sigma\left(\Delta \eta>0\right) + \sigma\left(\Delta \eta<0\right)}
\label{eq:fbasl}
\end{equation}
based on the pseudo-rapidity difference $\Delta\eta=\eta_{l^+}-\eta_{l^-}$.
The drawback of the leptonic asymmetry is that its value is much
smaller than the one based on the top quarks themselves.


At the LHC these forward--backward asymmetries cannot be measured, 
because of the symmetric $pp$ initial state. However, one can measure
the charge asymmetry using $\Delta \left|y\right| = \left|y_t\right|
-\left|y_{\bar{t}}\right|$:
\begin{equation}
A_{t\bar{t}}^{C}\;=\;\frac{\sigma\left(\Delta \left|y\right|>0\right) -  \sigma\left(\Delta \left|y\right|<0\right)}{ \sigma\left(\Delta \left|y\right|>0\right) + \sigma\left(\Delta \left|y\right|<0\right)}~.
\label{eq:cast}
\end{equation}
This asymmetry was measured by ATLAS \cite{ATLAS:2012an} and CMS \cite{Chatrchyan:2012cxa} and found to be in agreement with the Standard Model predictions. 
The leptonic charge asymmetry is defined analogously.

\begin{figure}[t!]
\centering
\includegraphics[width=0.7\textwidth]{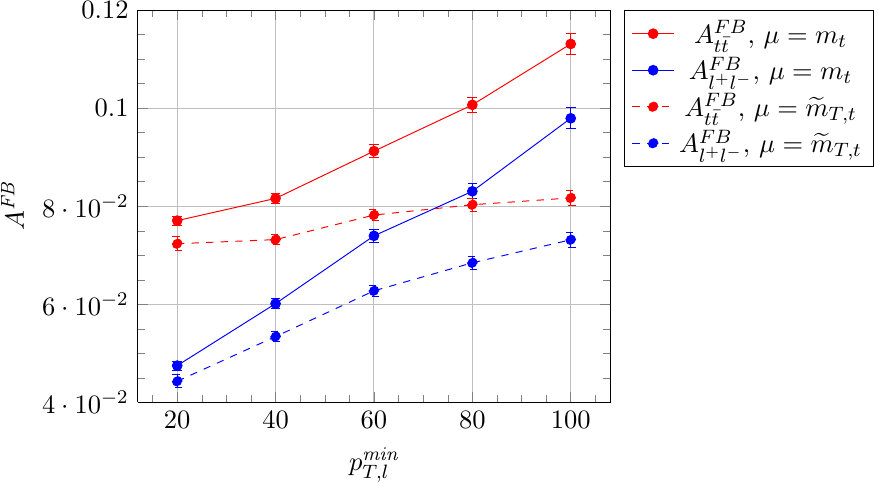}
\caption{Dependence of the $t\bar{t}$ and leptonic asymmetries on the
  minimal charged-lepton transverse momentum $p_{T,l}^{\mrm{min}}$.
  Note that the change of $A^{FB}$ with $p_{T,l}^{\rm{min}}$ also
  depends on the scale choice.}
\label{fig:asymptmin}
\end{figure}

Here we focus on a study of the forward--backward asymmetries in
$p\bar{p}$ collisions at $\sqrt{s}=1.96\mrm{\:TeV}$, where we apply
the anti-$k_T$ jet algorithm and the constraints $\Delta R>0.4$,
$p_{T,b}>20\gev$, $p_{T,l}>20\gev$, $|\eta_b|<2.5$, $|\eta_l|<2.5$,
$\slashed{p}_T>25\gev$.
As expected, we observe NLO corrections leading to a shift towards
positive values in both the $\Delta y$ and $\Delta\eta$ distributions.
We also investigated the dependence of the asymmetry on the $p_T$
requirement on the $l^\pm$ transverse momentum, $p_{T,l}^{\mrm{min}}$,
and on the scale choice, where we compared the scales
$\mu=\mu_\mrm{R}=\mu_\mrm{F}=\{m_t,\,\tilde m_{T,t}\}$ defining
$\tilde m_{T,t}=\sqrt{m_t^2+p^2_{T,\mrm{lead.jet}}}$.
The results are shown in Figure~\ref{fig:asymptmin}. We observe that
the correlation between $A_{t\bar{t}}^{FB}$ and $A_{l^+l^-}^{FB}$
becomes stronger as we increase $p_{T,l}^{\rm{min}}$, see also
Ref.~\cite{Falkowski:2012cu}. It also becomes clear that the change of
$A_{t\bar{t}}^{FB}$ and $A_{l^+l^-}^{FB}$ with $p_{T,l}^{\mrm{min}}$
depends rather strongly on the scale choice.
A more detailed study is given in Ref.~\cite{Heinrich:2013qaa}.


\section{Conclusions}

We have calculated the NLO QCD corrections to the production of a 
$W^+W^-$ pair in association with two $b$-jets, including the leptonic
decays of the $W$ bosons, and full off-shell effects of the top quarks.
We use our results to study the observable $m_{lb}$, the invariant mass of 
a charged lepton and a $b$-jet,  which is relevant for top quark mass
measurements.
We also studied the correlation between the $t\bar t$ and leptonic
forward--backward asymmetries at the Tevatron, and their dependence 
on the minimum value required for the charged lepton transverse
momentum, $p_{T,l}^{\mrm{min}}$, and on the scale choice.
 

\section*{Acknowledgements}

We would like to thank the \textsc{GoSam} collaboration, Stefan H\"oche, Richard Nisius and Andreas Maier for useful discussions.



\bibliographystyle{iopart-num} 
\providecommand{\newblock}{}


\begin{thebibliography}{10}
\expandafter\ifx\csname url\endcsname\relax
  \def\url#1{{\tt #1}}\fi
\expandafter\ifx\csname urlprefix\endcsname\relax\def\urlprefix{URL }\fi
\providecommand{\eprint}[2][]{\url{#2}}

\bibitem{Nason:1987xz}
Nason P, Dawson S and Ellis R~K 1988 {\em Nucl.Phys.\/} {\bf B303} 607

\bibitem{Nason:1989zy}
Nason P, Dawson S and Ellis R~K 1989 {\em Nucl.Phys.\/} {\bf B327} 49--92

\bibitem{Beenakker:1990maa}
Beenakker W, van Neerven W, Meng R, Schuler G and Smith J 1991 {\em
  Nucl.Phys.\/} {\bf B351} 507--560

\bibitem{Mangano:1991jk}
Mangano M~L, Nason P and Ridolfi G 1992 {\em Nucl.Phys.\/} {\bf B373} 295--345

\bibitem{Frixione:1995fj}
Frixione S, Mangano M~L, Nason P and Ridolfi G 1995 {\em Phys.Lett.\/} {\bf
  B351} 555--561 (\textit{Preprint} \eprint{hep-ph/9503213})

\bibitem{Czakon:2013goa}
Czakon M, Fiedler P and Mitov A 2013 {\em Phys.Rev.Lett.\/} {\bf 110} 252004
  (\textit{Preprint} \eprint{1303.6254})

\bibitem{Beenakker:1993yr}
Beenakker W, Denner A, Hollik W, Mertig R, Sack T {\em et~al.\/} 1994 {\em
  Nucl.Phys.\/} {\bf B411} 343--380

\bibitem{Melnikov:2009dn}
Melnikov K and Schulze M 2009 {\em JHEP\/} {\bf 0908} 049 (\textit{Preprint}
  \eprint{0907.3090})

\bibitem{Melnikov:2011ai}
Melnikov K and Schulze M 2011 {\em Phys.Lett.\/} {\bf B700} 17--20
  (\textit{Preprint} \eprint{1103.2122})

\bibitem{Biswas:2010sa}
Biswas S, Melnikov K and Schulze M 2010 {\em JHEP\/} {\bf 1008} 048
  (\textit{Preprint} \eprint{1006.0910})

\bibitem{Denner:2010jp}
Denner A, Dittmaier S, Kallweit S and Pozzorini S 2011 {\em Phys.Rev.Lett.\/}
  {\bf 106} 052001 (\textit{Preprint} \eprint{1012.3975})

\bibitem{Denner:2012yc}
Denner A, Dittmaier S, Kallweit S and Pozzorini S 2012 {\em JHEP\/} {\bf 1210}
  110 (\textit{Preprint} \eprint{1207.5018})

\bibitem{Bevilacqua:2010qb}
Bevilacqua G, Czakon M, van Hameren A, Papadopoulos C~G and Worek M 2011 {\em
  JHEP\/} {\bf 1102} 083 (\textit{Preprint} \eprint{1012.4230})

\bibitem{Heinrich:2013qaa}
Heinrich G, Maier A, Nisius R, Schlenk J and Winter J 2013  (\textit{Preprint}
  \eprint{1312.6659})

\bibitem{Frederix:2013gra}
Frederix R 2013  (\textit{Preprint} \eprint{1311.4893})

\bibitem{Cascioli:2013wga}
Cascioli F, Kallweit S, Maierhöfer P and Pozzorini S 2013  (\textit{Preprint}
  \eprint{1312.0546})

\bibitem{Cullen:2011ac}
Cullen G, Greiner N, Heinrich G, Luisoni G, Mastrolia P {\em et~al.\/} 2012
  {\em Eur.Phys.J.\/} {\bf C72} 1889 (\textit{Preprint} \eprint{1111.2034})

\bibitem{Ossola:2006us}
Ossola G, Papadopoulos C~G and Pittau R 2007 {\em Nucl.Phys.\/} {\bf B763}
  147--169 (\textit{Preprint} \eprint{hep-ph/0609007})

\bibitem{Ellis:2007br}
Ellis R, Giele W and Kunszt Z 2008 {\em JHEP\/} {\bf 0803} 003
  (\textit{Preprint} \eprint{0708.2398})

\bibitem{Mastrolia:2010nb}
Mastrolia P, Ossola G, Reiter T and Tramontano F 2010 {\em JHEP\/} {\bf 1008}
  080 (\textit{Preprint} \eprint{1006.0710})

\bibitem{Heinrich:2010ax}
Heinrich G, Ossola G, Reiter T and Tramontano F 2010 {\em JHEP\/} {\bf 1010}
  105 (\textit{Preprint} \eprint{1008.2441})

\bibitem{vanDeurzen:2013pja}
van Deurzen H 2013 {\em Acta Phys.Polon.\/} {\bf B44} 2223--2230

\bibitem{vanDeurzen:2013saa}
van Deurzen H, Luisoni G, Mastrolia P, Mirabella E, Ossola G {\em et~al.\/}
  2013  (\textit{Preprint} \eprint{1312.6678})

\bibitem{Binoth:2005ff}
Binoth T, Guillet J~P, Heinrich G, Pilon E and Schubert C 2005 {\em JHEP\/}
  {\bf 0510} 015 (\textit{Preprint} \eprint{hep-ph/0504267})

\bibitem{Binoth:2008uq}
Binoth T, Guillet J~P, Heinrich G, Pilon E and Reiter T 2009 {\em
  Comput.Phys.Commun.\/} {\bf 180} 2317--2330 (\textit{Preprint}
  \eprint{0810.0992})

\bibitem{Cullen:2011kv}
Cullen G, Guillet J~P, Heinrich G, Kleinschmidt T, Pilon E {\em et~al.\/} 2011
  {\em Comput.Phys.Commun.\/} {\bf 182} 2276--2284 (\textit{Preprint}
  \eprint{1101.5595})

\bibitem{Guillet:2013msa}
Guillet J~P, Heinrich G and von Soden-Fraunhofen J 2013  (\textit{Preprint}
  \eprint{1312.3887})

\bibitem{vanHameren:2010cp}
van Hameren A 2011 {\em Comput.Phys.Commun.\/} {\bf 182} 2427--2438
  (\textit{Preprint} \eprint{1007.4716})

\bibitem{Gleisberg:2007md}
Gleisberg T and Krauss F 2008 {\em Eur.Phys.J.\/} {\bf C53} 501--523
  (\textit{Preprint} \eprint{0709.2881})

\bibitem{Gleisberg:2008ta}
Gleisberg T, Hoeche S, Krauss F, Schonherr M, Schumann S {\em et~al.\/} 2009
  {\em JHEP\/} {\bf 0902} 007 (\textit{Preprint} \eprint{0811.4622})

\bibitem{Binoth:2010xt}
Binoth T, Boudjema F, Dissertori G, Lazopoulos A, Denner A {\em et~al.\/} 2010
  {\em Comput.Phys.Commun.\/} {\bf 181} 1612--1622 (\textit{Preprint}
  \eprint{1001.1307})

\bibitem{Alioli:2013nda}
Alioli S, Badger S, Bellm J, Biedermann B, Boudjema F {\em et~al.\/} 2014 {\em
  Comput.Phys.Commun.\/} {\bf 185} 560--571 (\textit{Preprint}
  \eprint{1308.3462})

\bibitem{gosamtalks1}
Luisoni G {\em In these proceedings.\/}

\bibitem{gosamtalks2}
Mastrolia P {\em In these proceedings.\/}

\bibitem{Denner:2006ic}
Denner A and Dittmaier S 2006 {\em Nucl.Phys.Proc.Suppl.\/} {\bf 160} 22--26
  (\textit{Preprint} \eprint{hep-ph/0605312})

\bibitem{Nagy:1998bb}
Nagy Z and Trocsanyi Z 1999 {\em Phys.Rev.\/} {\bf D59} 014020
  (\textit{Preprint} \eprint{hep-ph/9806317})

\bibitem{AlcarazMaestre:2012vp}
Alcaraz~Maestre J {\em et~al.\/} 2012  (\textit{Preprint} \eprint{1203.6803})

\bibitem{Martin:2009iq}
Martin A, Stirling W, Thorne R and Watt G 2009 {\em Eur.Phys.J.\/} {\bf C63}
  189--285 (\textit{Preprint} \eprint{0901.0002})

\bibitem{Cacciari:2005hq}
Cacciari M and Salam G~P 2006 {\em Phys.Lett.\/} {\bf B641} 57--61
  (\textit{Preprint} \eprint{hep-ph/0512210})

\bibitem{Cacciari:2008gp}
Cacciari M, Salam G~P and Soyez G 2008 {\em JHEP\/} {\bf 0804} 063
  (\textit{Preprint} \eprint{0802.1189})

\bibitem{Cacciari:2011ma}
Cacciari M, Salam G~P and Soyez G 2012 {\em Eur.Phys.J.\/} {\bf C72} 1896
  (\textit{Preprint} \eprint{1111.6097})

\bibitem{Chetyrkin:1999ys}
Chetyrkin K and Steinhauser M 1999 {\em Phys.Rev.Lett.\/} {\bf 83} 4001--4004
  (\textit{Preprint} \eprint{hep-ph/9907509})

\bibitem{Melnikov:2000qh}
Melnikov K and Ritbergen T~v 2000 {\em Phys.Lett.\/} {\bf B482} 99--108
  (\textit{Preprint} \eprint{hep-ph/9912391})

\bibitem{Agashe:2013hma}
Agashe K {\em et~al.\/} 2013  (\textit{Preprint} \eprint{1311.2028})

\bibitem{Juste:2013dsa}
Juste A, Mantry S, Mitov A, Penin A, Skands P {\em et~al.\/} 2013
  (\textit{Preprint} \eprint{1310.0799})

\bibitem{ATLAS:2012aj}
Aad G {\em et~al.\/} (ATLAS) 2012 {\em Eur.Phys.J.\/} {\bf C72} 2046
  (\textit{Preprint} \eprint{1203.5755})

\bibitem{ATLAS-CONF-2013-077}
 2013 {Measurement of the Top Quark Mass in Dileptonic Top Quark Pair Decays
  with $\sqrt{s}=7$ TeV ATLAS Data} Tech. Rep. ATLAS-CONF-2013-077 CERN Geneva

\bibitem{Chatrchyan:2012ea}
Chatrchyan S {\em et~al.\/} (CMS) 2012 {\em Eur.Phys.J.\/} {\bf C72} 2202
  (\textit{Preprint} \eprint{1209.2393})

\bibitem{Aaltonen:2011dr}
Aaltonen T {\em et~al.\/} (CDF) 2011 {\em Phys.Rev.\/} {\bf D83} 111101
  (\textit{Preprint} \eprint{1105.0192})

\bibitem{Abazov:2012rp}
Abazov V~M {\em et~al.\/} (D0) 2012 {\em Phys.Rev.\/} {\bf D86} 051103
  (\textit{Preprint} \eprint{1201.5172})

\bibitem{Abazov:2011rq}
Abazov V~M {\em et~al.\/} (D0) 2011 {\em Phys.Rev.\/} {\bf D84} 112005
  (\textit{Preprint} \eprint{1107.4995})

\bibitem{Aaltonen:2011kc}
Aaltonen T {\em et~al.\/} (CDF) 2011 {\em Phys.Rev.\/} {\bf D83} 112003
  (\textit{Preprint} \eprint{1101.0034})

\bibitem{Kuhn:2011ri}
Kuhn J~H and Rodrigo G 2012 {\em JHEP\/} {\bf 1201} 063 (\textit{Preprint}
  \eprint{1109.6830})

\bibitem{Hollik:2011ps}
Hollik W and Pagani D 2011 {\em Phys.Rev.\/} {\bf D84} 093003
  (\textit{Preprint} \eprint{1107.2606})

\bibitem{Bernreuther:2012sx}
Bernreuther W and Si Z~G 2012 {\em Phys.Rev.\/} {\bf D86} 034026
  (\textit{Preprint} \eprint{1205.6580})

\bibitem{ATLAS:2012an}
Aad G {\em et~al.\/} (ATLAS) 2012 {\em Eur.Phys.J.\/} {\bf C72} 2039
  (\textit{Preprint} \eprint{1203.4211})

\bibitem{Chatrchyan:2012cxa}
Chatrchyan S {\em et~al.\/} (CMS) 2012 {\em Phys.Lett.\/} {\bf B717} 129--150
  (\textit{Preprint} \eprint{1207.0065})

\bibitem{Falkowski:2012cu}
Falkowski A, Mangano M~L, Martin A, Perez G and Winter J 2013 {\em Phys.Rev.\/}
  {\bf D87} 034039 (\textit{Preprint} \eprint{1212.4003})

\end{thebibliography}

\end{document}